\documentclass[superscriptaddress,showpacs,amssymb,10pt,aps,prd,reprint,longbibliography,footinbib]{revtex4-1}

\usepackage{graphicx,epsfig,amssymb,times} 
\usepackage{amsmath,amsfonts}
\usepackage{bm}
\usepackage{epstopdf}
\usepackage[linktocpage,colorlinks]{hyperref}
\usepackage[caption=false]{subfig}
\usepackage[usenames]{color}     
\usepackage{float}
\usepackage{natbib}

\definecolor{coolblack}{rgb}{0.0, 0.18, 0.39}
\definecolor{darkred}{rgb}{0.5,0,0}
\definecolor{darkgreen}{rgb}{0,0.5,0}
\definecolor{darkblue}{rgb}{0,0,0.5}
\definecolor{lapislazuli}{rgb}{0.15, 0.38, 0.61}
\definecolor{venetianred}{rgb}{0.78, 0.03, 0.08}
\definecolor{bleudefrance}{rgb}{0.19, 0.55, 0.91}
\definecolor{dogwoodrose}{rgb}{0.84, 0.09, 0.41}
\hypersetup{colorlinks=true, citecolor=darkblue, linkcolor=darkblue, 
urlcolor = darkblue}

\def\be{\begin{equation}}
\def\ee{\end{equation}}

\newcommand{\bea}{\begin{eqnarray}}
\newcommand{\eea}{\end{eqnarray}}
\newcommand{\ben}{\begin{enumerate}}
\newcommand{\een}{\end{enumerate}}
\newcommand{\bi}{\begin{itemize}}
\newcommand{\ei}{\end{itemize}}

\newcommand{\rs}{R_{\rm S}}
\newcommand{\mh}{M_{\rm BH}}

\def\ga{\mathrel{\raise.3ex\hbox{$>$\kern-.75em\lower1ex\hbox{$\sim$}}}}
\def\la{\mathrel{\raise.3ex\hbox{$<$\kern-.75em\lower1ex\hbox{$\sim$}}}}

\def\l{\left}
\def\r{\right}
\def\be{\begin{equation}}
\def\ee{\end{equation}}

\def\I_M{{I_{\scriptscriptstyle M\times M}}}

\def\be{\begin{equation}}
\def\ee{\end{equation}}
\def\bea{\begin{eqnarray}}
\def\eea{\end{eqnarray}}
\newcommand{\beq}{\begin{eqnarray}}
\newcommand{\eeq}{\end{eqnarray}}
\def\pa{\partial}

\newcommand{\beqal}{\begin{eqnarray}\label}
\newcommand{\beqa}{\begin{eqnarray}}
\newcommand{\eeqa}{\end{eqnarray}}

\begin{document}

\title{\large Black holes with surrounding matter and rainbow scattering}

\author{Luiz C. S. Leite}\email{luizcsleite@ufpa.br}
\affiliation{Faculdade de F\'{\i}sica, Universidade 
Federal do Par\'a, 66075-110, Bel\'em, Par\'a, Brazil}

\author{Caio F. B. Macedo}\email{caiomacedo@ufpa.br}
\affiliation{Campus Salin\'opolis, Universidade Federal do Par\'a,
68721-000, Salin\'opolis, Par\'a, Brazil}

\author{Lu\'is C. B. Crispino}\email{crispino@ufpa.br}
\affiliation{Faculdade de F\'{\i}sica, Universidade 
Federal do Par\'a, 66075-110, Bel\'em, Par\'a, Brazil}

\date{\today}


\begin{abstract}
The scattering of light by water droplets can produce one of the most beautiful phenomena in nature: the rainbow. This optical phenomenon has analogues in molecular, atomic, and nuclear physics. Recently, rainbow scattering has been shown to arise from the gravitational interaction of a scalar field with a compact horizonless object. We show that rainbow scattering can also occur in the background of a black hole with surrounding matter. We study the scattering of null geodesics and planar massless scalar waves by Schwarzschild black holes surrounded by a thin spherical shell of matter. We explore various configurations of this system, analyzing changes in mass fraction and radius of the shell. We show that the deflection function can present stationary points, which leads to rainbow scattering. We analyze the large-angle scattered amplitude as a function of the shell's parameters, showing that it presents a nonmonotonic behavior. 
\end{abstract}

\maketitle

\section{Introduction}

Scattering of waves and particles is a very important tool in many branches of physics. Most of our knowledge about atomic and nuclear physics comes from scattering experiments as, for instance, the discovery of the Higgs boson in the ATLAS and CMS experiments at CERN's Large Hadron Collider~\cite{Higgsboson2012}. Moreover, the scattering of light by the Sun provided the first experimental test of general relativity (GR) about a hundred years ago~\cite{davidson1920ix}.

Black holes (BHs) scatter radiation propagating in their vicinities, a subject which has been widely explored (see, e.g., Ref.~\cite{Futterman:1988ni}). In the standard scenario, one considers that a monochromatic wave (propagating from infinity) impinges upon a BH and is scattered due to the gravitational interaction with the fixed background. Using this approach, many authors have studied the time-independent scattering of planar waves by BHs~(see, e.g., Refs.~\cite{Matzner:1968,Vishveshwara:1970zz,PhysRevD.7.2807,PhysRevD.10.1059,Fabbri:1975sa,Sanchez:1977vz,Glampedakis2001,Dolan:2006vj,Dolan:2008kf,Crispino:2009xt,Crispino_2009:prd79_064022,Chen2013,PhysRevD.90.064027,Macedo:2015qma,PhysRevD.92.084056,cotuaescu2016partial,cotuaescu2016partialrn,Sporea2017}). However, most of the attention in the literature has been drawn to the study of wave scattering by isolated BH (IBH) solutions~\footnote{By isolated, we mean electrovacuum solutions.}. 

When considering astrophysical setups, BHs are likely to be surrounded by matter, typically accretion disks~\cite{Narayan:2005ie}. Therefore, the study of the scattering of fields by BHs in the presence of surrounding matter is very appropiate. Taking this into account, we study the scattering of the massless scalar field by a Schwarzschild BH surrounded by a thin spherical shell of matter~\cite{Shell}, which is usually dubbed a {\it{dirty black hole}} (DBH) and represents a simplified version of a real astrophysical setup. DBH configurations have been used in the literature to seek the influence of matter surrounding BHs---for instance, in gravitational-wave astronomy~\cite{Barausse}, in BH quasinormal modes~\cite{Leung:1999rh,Medved2004}, and in BH absorption~\cite{Macedo2016}. Recently, a configuration composed by a traversable wormhole with a thick shell of matter at a distance from its throat has been used to study the influence of the astrophysical environment in the echoes of the surface of the compact objects~\cite{Konoplya:2018yrp}.  

It is also interesting to compare the scattering patterns of IBHs and DBHs with those of configurations without event horizons. In Ref.~\cite{Dolan:2017rtj}, it has been shown that the scattering by compact stellar configurations can result in interesting phenomena, closely related to optics. One of them, the rainbow scattering~\cite{Nussenzveig:79,Nussenzveigdoi:10.1063/1.1664747}, appears whenever there is an extremum point in the deflection angle, as a function of the impact parameter, enhancing the scattering amplitude at that angle. 

We show that the existence of matter surrounding the BH gives support to rainbow scattering, a feature that is not present in the case of an IBH. We arrive at this conclusion by investigating the scattering of null geodesics and planar massless scalar waves by DBHs. In Sec.~\ref{sec:metric}, we describe the DBH configuration used in our study and review some features of the massless scalar wave propagation in the DBH spacetime, presenting the boundary conditions suitable for the scattering problem. In Sec.~\ref{sec:scattering}, we present expressions of the differential scattering cross sections for null geodesics and scalar waves, and we briefly discuss the semiclassical approximation for the glory scattering. We also show that the deflection function possesses stationary points for some DBH configurations. In Sec.~\ref{sec:num_results}, we describe the numerical methods and present our numerical results. We present some final remarks in Sec.~\ref{sec:final_remarks}. Throughout this paper, natural units ($c=G=\hslash=1$) are used.

\section{Spacetime and wave propagation}\label{sec:metric}
We shall focus on null geodesics and scalar waves impinging into a Schwarzschild BH surrounded by a thin spherical shell. The corresponding spacetime line element can be written as
\be
ds^2=-A(r)dt^2+\frac{1}{1-2\mu(r)/r}dr^2+r^2 d\Omega^2,\label{eq:sphericalspacetime}
\ee
where $d\Omega^2$ is the solid angle element of the two-sphere, and $A(r)$ and $\mu(r)$ are radial dependent functions, with $\mu(r)$ being the mass function. Considering a spherical shell composed by a perfect fluid with a radius fixed at $\rs$, we have that in the outer region of the configuration composed by the Schwarzschild BH with the spherical shell, $r>\rs$, the line element reads
\be
ds^2=-\l(1-\frac{2M}{r}\r)dt^2+\frac{1}{1-2M/r}dr^2+r^2 d\Omega^2,\label{eq:outerspacetime}
\ee
with $M$ being the total Arnowitt-Deser-Misner (ADM) mass. In the region $2\mh<r<\rs$, by demanding the metric function $A(r)$ to be continuous across the spherical shell, we obtain
\be
ds^2=-\alpha\l(1-\frac{2\mh}{r}\r)dt^2+\frac{1}{1-2\mh/r}dr^2+r^2 d\Omega^2,\label{eq:innerspacetime}
\ee
with $\alpha$ being given by
\be
\alpha=\frac{1-2M/\rs}{1-2\mh/\rs},
\ee
where $\mh$ is the BH's relative mass. Note that both the BH's and the shell's masses are taken into account in the total ADM mass. 

Restrictions on the shell position and mass can be imposed by requiring the shell's matter to obey specific energy conditions. By imposing the dominant energy condition (DEC) or the strong energy condition (SEC), restrictions are obtained for the minimum possible value of the shell radius (see Ref.~\cite{Macedo2016} for explicit expressions of the minimum shell radius related to each energy condition).

Massless scalar waves are described by the Klein-Gordon equation, which can be written as follows:
\be
\pa_a(g^{ab}\sqrt{-g}\pa_b\Phi)=0,
\ee
with $\pa_a\equiv\pa/\pa x^a$.

The massless scalar field~$\Phi$ can be analyzed using separation of variables, namely
\be
\Phi=\frac{\phi(r)}{r}Y_{lm}e^{-i\omega t},
\label{eq:decom}
\ee
where $Y_{lm}$ are the spherical harmonics. We are left with the following radial equation for $\phi$:
\be
\frac{d^2\phi}{dx^2}+\l[\omega^2-V(x)\r]\phi=0,
\label{eq:kgr}
\ee
where the effective potential is given by
\be
V(x)=A\l[\frac{l(l+1)}{r^2}+\frac{2\mu}{r^3}\r],
\ee
and $x$ is a radial-like coordinate related to $r$ via
\be
dx=\frac{dr}{\sqrt{A(1-2\mu/r)}}.\label{eq:tortoise_coord}
\ee
It follows directly from Eq.~\eqref{eq:tortoise_coord} that the BH horizon is located at $x\to-\infty$, and the spatial infinity at $x\to+\infty$.

We are interested in solutions of Eq.~\eqref{eq:kgr} that satisfy the following boundary conditions:
\be
\phi(x)\sim\l\{
\begin{array}{ll}
	{\cal A}_{\omega l }e^{-i \omega x}+{\cal R}_{\omega l }e^{i \omega x},&x\to+\infty, \\
	{\cal T}_{\omega l }e^{-i\omega x},& x\to-\infty,
\end{array}\r.\label{eq:inmodes}
\ee
with the coefficients related by
\be
|{\cal A}_{\omega l }|^2=|{\cal T}_{\omega l }|^2+|{\cal R}_{\omega l }|^2.
\ee

\section{Scattering cross section} \label{sec:scattering}
Following Ref.~\cite{Futterman:1988ni}, we write the differential scattering cross section $\frac{d\sigma}{d\Omega}$ as a partial wave series, namely
\be
\frac{d\sigma}{d\Omega}=\l|\frac{1}{2i\omega}\sum_{l=0}^{\infty}(2l+1)(e^{2i \delta_{\omega l}}-1)P_l(\cos\theta)\r|^2,
\label{eq:scattering}
\ee
where $\theta$ is the scattering angle defined in the interval $[0,\,\pi]$ and the phase shifts $\delta_{\omega l}$ read
\be
e^{2i \delta_{\omega l}}=(-1)^{l+1}\frac{{\cal R}_{\omega l}}{{\cal A}_{\omega l}}.
\ee

One can show that, for small scattering angles ($\theta\rightarrow0$), the differential scattering cross section diverges as~\cite{sanchez1977wave}
\be
\frac{d\sigma}{d\Omega}\sim 16\frac{\mh^2}{\theta^4}.
\ee
We recall that the corresponding classical scattering cross section also presents the same behavior for small scattering angles~\cite{Collins:1973xf}, as a consequence of the long-range feature of the gravitational interaction~\cite{Sanchez:1977vz}.

\subsection{Geodesic scattering}

We can analyze the classical limit of the differential scattering cross section through a geodesic analysis. For the case of a Schwarzschild BH surrounded by a thin spherical shell, the analysis of Ref.~\cite{Collins:1973xf} can be applied. The differential scattering cross section for null geodesics is given by
\be
\frac{d\sigma}{d\Omega}=\frac{b}{\sin\theta}\l|\frac{d\chi(b)}{db}\r|^{-1},
\label{eq:scatteringgeo}
\ee
where $b$ is the impact parameter of the scattered null geodesic, and the deflection angle $\chi$ is related to the scattering angle $\theta$ by $\chi=\pm\theta-2n\pi$, with $n=0$, $1$, $2$,$\ldots$.

The impact parameter as a function of the scattering angle can be found by analyzing null geodesics incoming from infinity. Without loss of generality, due to the spherical symmetry, we can restrict ourselves to the motion in the equatorial plane. The analysis is similar to the one presented in Ref.~\cite{Chandrasekhar:1985kt}. We obtain
\be
\l(\frac{du}{d\varphi}\r)^2=\l(1-\frac{2 \mu}{r}\r) \frac{1-A b^2 u^2}{A b^2 },
\label{eq:nullgeo}
\ee
where $u\equiv1/r$, and we can write the impact parameter as~$b=L/E$, with $L$ and $E$ being the angular momentum and the energy, respectively. Null circular geodesics---also known as light rings ---can be found by solving Eq.~\eqref{eq:nullgeo}, imposing $du/d\varphi=0$ and $d^2u/d\varphi^2=0$, obtaining the radius of the light ring $r_l$, and the corresponding impact parameter $b_l$. For an isolated Schwarzschild BH, for instance, there is only one light ring, located at $r_l=3M$, with $b_l=3\sqrt{3}M$. In the case of a BH surrounded by a spherical shell, we have three possible situations~\cite{Macedo2016}:

$(i)$ When $\rs<3\mh$, we have
\be
r_{l}=3M,~	 b_{l}=3\sqrt{3}M.
\ee
$(ii)$ For $3\mh<\rs<3M$, we have two light rings, namely
\be
r_{l_-}= 3\mh,~{\rm with}~b_{l_-}=3\sqrt{3}\mh/\sqrt{\alpha},
\ee
and
\be
r_{l_+}=3M,~{\rm with}~b_{l_+}=3\sqrt{3}M.
\ee
$(iii)$ When $\rs>3M$, we have
\be
r_{l}=3\mh,~{\rm with}~b_{l}=3\sqrt{3}\mh/\sqrt{\alpha}.
\ee

The critical impact parameter $b_c$ is closely related to the impact parameter of circular null geodesics $b_l$. This is due to the fact that circular null geodesics are unstable, representing a maximum point in the effective potential~\cite{Chandrasekhar:1985kt}. For $b>b_c$ a null geodesic is scattered. For the cases $\rs<3\mh$ and $\rs>3M$, we have $b_c=b_l$. In the case with two light rings, we have
\be
b_c=\min(b_{l_+},b_{l_-}).
\ee

By integrating directly Eq.~\eqref{eq:nullgeo}, we obtain the deflection angle as a function of the impact parameter through~\cite{Collins:1973xf}
\be
\chi(b)=2\beta(b)	-\pi,
\label{eq:deflect}
\ee
with
\be
\beta=\int_0^{u_0}du\l[\l(1-\frac{2 \mu}{r}\r) \frac{1-A b^2 u^2}{A b^2 }\r]^{-1/2},
\ee
where $u_0=1/r_0$ denotes the turning point.

By inverting Eq.~\eqref{eq:deflect}, writing $b(\chi)$, we can use Eq.~\eqref{eq:scattering} to obtain the differential scattering cross section for null geodesics.

Many interesting features appear in the discussion of the scattering of null geodesics (see, e.g., Ref.~\cite{Collins:1973xf}). The classical differential scattering cross section, given by Eq.~\eqref{eq:scatteringgeo}, formally diverges for $\theta\approx\pi$, which is related to the glory effect~\cite{Ford1959AnPhy...7..259F,newton1982scattering,nussenzveig2006diffraction}, an enhancement of the scattering amplitude for large scattering angles (see also Sec.~\ref{sec:glory}). There is also a divergence at scattering angles for which $d\chi/db=0$, which in optics is associated to rainbows, and therefore scattering phenomena near those angles receive the name \textit{rainbow scattering} (for an example in a compact object background, see Ref.~\cite{Dolan:2017rtj})~\cite{Ford1959AnPhy...7..259F,newton1982scattering,nussenzveig2006diffraction}. The rainbow scattering angle is given by $\theta_r\equiv|\chi(b_r)|$, and the rainbow impact parameter $b_r$ is obtained from the condition $\l.\frac{d\chi}{db}\r|_{b=b_r}=0$. The deflection angle is divergent for a critical value of the impact parameter $b=b_c$. In this case, the particle orbits around the compact object an infinite number of times, characterizing a  phenomenon called \textit{orbiting}, and this kind of scattering is known as \textit{spiral scattering}~\cite{Ford1959AnPhy...7..259F,newton1982scattering,nussenzveig2006diffraction}.
 
\begin{figure}[h!]%
\includegraphics[width=\columnwidth]{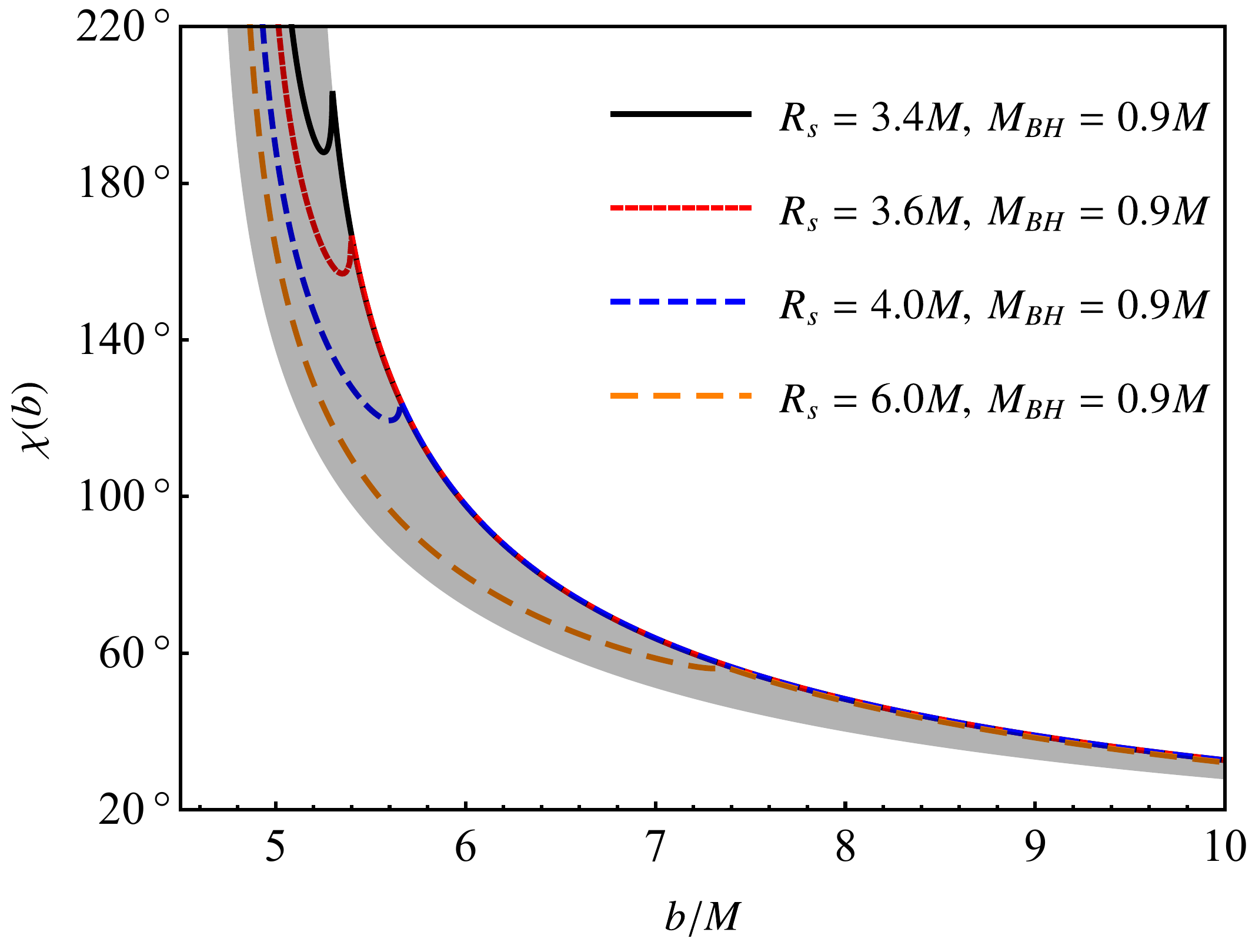}
\caption{Deflection angle as a function of the impact parameter for $\mh=0.9M$, and shell positions $\rs/M=6.0$, $4.0$, $3.6$ and $3.4$, with the latter case being represented in Fig.~\ref{fig:geoscat}. The shaded region is delimited by the isolated BH cases, with $\mh=0.9M$ (from below) and $\mh=M$ (from above). In all cases, the deflection function diverges at the respective critical impact parameter~$b=b_{\rm{c}}$.
}%
\label{fig:geodefleang}%
\end{figure}

In Fig.~\ref{fig:geodefleang}, we show a comparison between the deflection angle $\chi$ of an isolated Schwarzschild BH~(ISBH) and DBHs. The shaded region is delimited by ISBH cases: from below for $\mh=0.9M$, and from above for $\mh=M$. We note that, for the ISBH, the deflection angle is a monotonic function. For the case of a DBH, there may show up some extremal points~($d\chi/db=0$), depending on the set of parameters $(\rs,\mh)$. Rainbow scattering was previously reported to exist in horizonless compact objects~\cite{Dolan:2017rtj}, and Fig.~\ref{fig:geodefleang} shows that it can occur for DBHs as well. We point out the presence of two extrema in the DBH case, in contrast with just one in the compact star configurations studied in Ref.~\cite{Dolan:2017rtj}. When stationary points are present in the deflection angle, i.e., $\l.\frac{d\chi}{db}\r|_{b=b_r}=0$, we shall refer to them as the maximum rainbow scattering angle $\theta_{r,max}\equiv|\chi(b_{r,max})|$ and the minimum rainbow scattering angle $\theta_{r,min}\equiv|\chi(b_{r,min})|$, which are related to a local maximum $\l(\l.\frac{d^2\chi}{db^2}\r|_{b=b_{r,max}}<0\r)$ and a local minimum $\l(\l.\frac{d^2\chi}{db^2}\r|_{b=b_{r,min}}>0\r)$ of the deflection function, respectively.

\begin{figure}[h!]%
\includegraphics[width=\columnwidth]{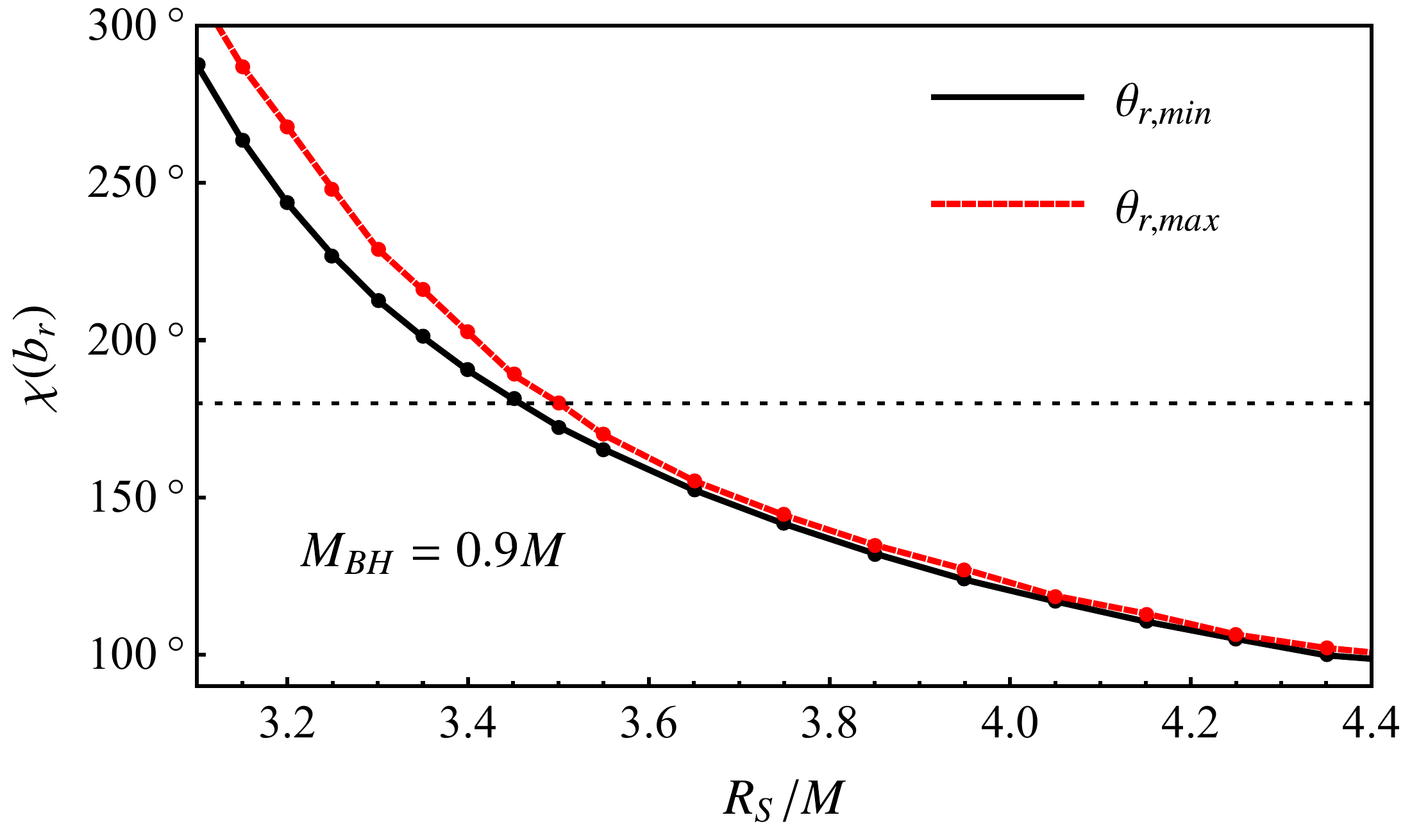}%
\caption{Rainbow scattering angles~($\theta_{r,min}$ and $\theta_{r,max}$) as  functions of the shell position, for $\mh=0.9 M$. The dark (black) line is a track of the relative minimum shown in Fig.~\ref{fig:geodefleang}, while the light (red) line is a track of the relative maximum. As $\rs$ increases, the two lines come together. The horizontal dotted line corresponds to $\chi(b_r)=\pi$.}%
\label{fig:rainbowangle}%
\end{figure}

In Fig.~\ref{fig:rainbowangle}, we show the rainbow scattering angle $\chi(b_r)$, as a function of the shell position, for the case $\mh=0.9M$. The two curves plotted in Fig.~\ref{fig:rainbowangle} are more separated~(and the rainbow angles larger) when the shell is nearer to the BH, getting closer to each other~(and the rainbow angles smaller) as the shell radius increases, eventually making the rainbow angles disappear (the Schwarzschild limiting case presents no rainbow angle). One interesting fact, which we shall revisit in Sec.~\ref{sec:glory}, is that when the shell is located at $\rs=3.45M$, the minimum rainbow angle, $\theta_{r,min}$, occurs approximately in the backward direction, $\theta_{r,min}\approx\pi$ (cf. the dashed horizontal line in Fig.~\ref{fig:rainbowangle}). Similarly, when the shell is located near $\rs\approx3.5M$, the maximum rainbow angle is also close to the antipodal direction, $\theta_{r,max}\approx\pi$~(cf. Fig.~\ref{fig:rainbowangle}).

\begin{figure}[h!]%
\includegraphics[width=\columnwidth]{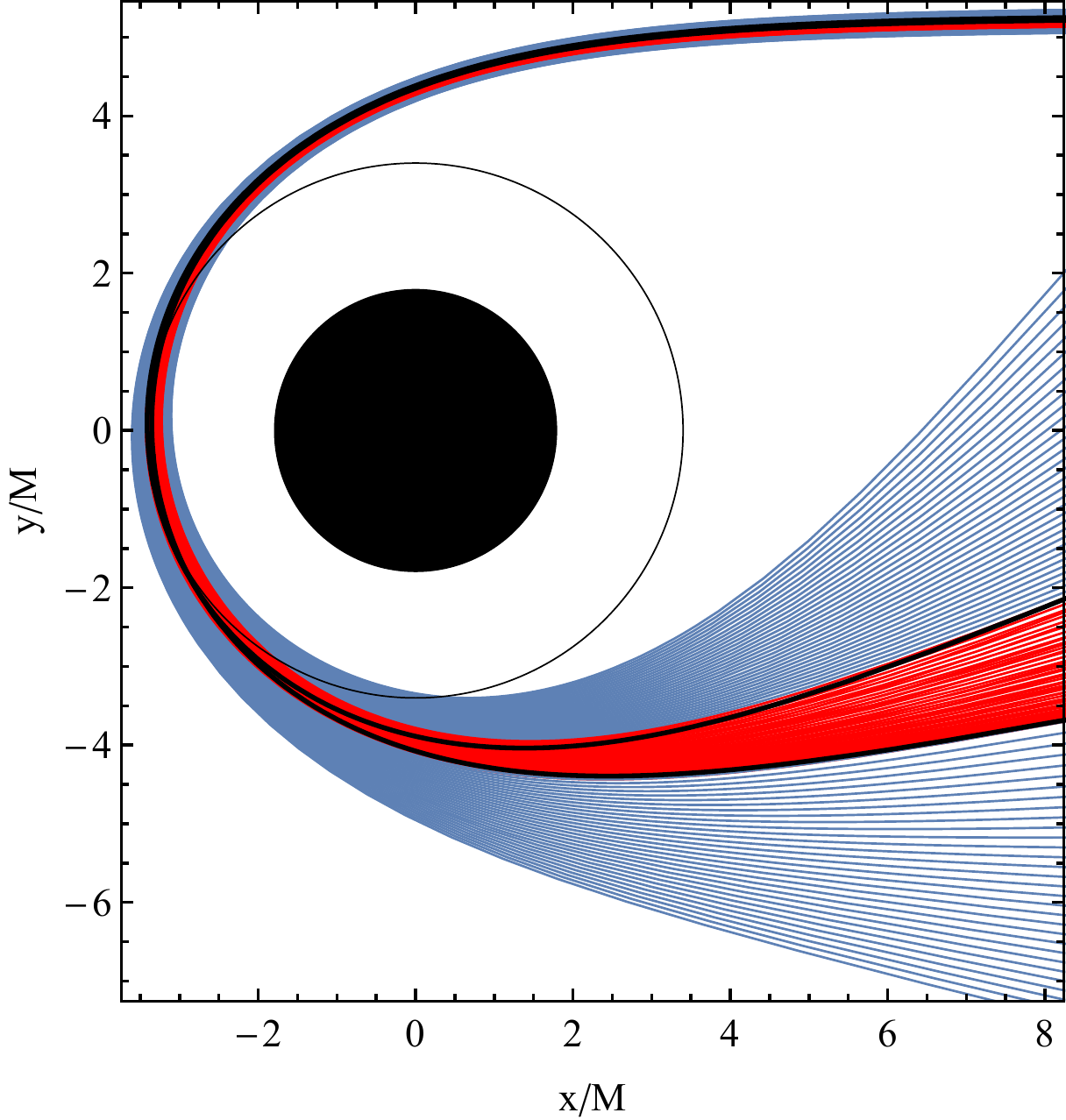}%
\caption{Scattering of null geodesics by a DBH. In this case, we have $\mh=0.9M$ and $\rs=3.4M$, such that we have extrema in the deflection angle~(cf. Fig.~\ref{fig:geodefleang}). We choose different impact parameters to illustrate the rainbow scattering. We choose incident geodesics with impact parameters equally spaced. Note that between the two bold (black) solid lines there is a concentration of lines, illustrating that in this region there are at least two geodesics with the same deflection angle. The (black) solid circle represents the location of the shell at $\rs=3.4M$.}%
\label{fig:geoscat}%
\end{figure}

In Fig.~\ref{fig:geoscat}, we exhibit a stream of null geodesics coming from infinity with impact parameters varying from $b=1.02b_c$ to $1.08b_c$, with a fixed step size of $0.0005b_c$. This configuration represents the case in which $\mh=0.9M$ and $\rs=3.4M$. For this case, the two extrema of the deflection angle are at $b\approx 1.053b_c$ (maximum) and $b\approx 1.06b_c$ (minimum). When $1.036<b/b_c<1.065$ (interval delimited by the two bold scattering trajectories in Fig.~\ref{fig:geoscat}) there are at least two geodesics with the same deflection angle, due to its nonmonotonical behavior. This characteristic leads to an amplification in the scattering amplitude at the rainbow angles, which, for this case, occurs for large scattering angles (cf. Fig.~\ref{fig:geodefleang}). Additionally, as in the case of horizonless compact objects~\cite{Dolan:2017rtj}, a caustic forms (cf. Fig.~\ref{fig:geoscat}) related to the rainbow angles.

\subsection{Semiclassical glory scattering: Wave approach for large angles}\label{sec:glory}

Another important approximation for the scattering of planar waves by BHs is the semiclassical approach to the glory scattering, valid for $\theta\approx \pi$. This well-known phenomenon in optics is also present in BH scattering, being reported in many previous works, occurring mainly because of the strong deflections close to the critical impact parameter~\cite{Matzner.31.1869}. In the geodesic approximation analysis, we have mentioned that the scattering cross section is singular in the limit $\theta\to\pi$. However, we do not have such a divergence in the partial wave results
, obtaining instead finite values for the scattering cross section at $\theta=\pi$. 

The semiclassical approach gives an analytical formula for the glory scattering, to which we can compare the full numerical results from the partial wave analysis (see also Ref.~\cite{Futterman:1988ni} for more details). The semiclassical glory approximation can be written as~\cite{Matzner.31.1869}
\be
\l.\frac{d\sigma}{d\Omega}\r|_{\theta\approx \pi}=2\pi\omega {b_g^2}\l|\frac{db}{d\chi}\r|_{\chi= \pi}J_0^2(\omega b_g\sin\theta)
\label{eq:semiglory}
\ee
where $J_0$ is the Bessel function of the first kind (of order 0), and $b_g \equiv b(\chi=\pi)$. In order to improve the semiclassical approximation, one can consider the contributions from the orbiting cases, as described in Refs.~\cite{Crispino_2009:prd79_064022,Macedo:2015qma}.

One interesting feature that follows from Eq.~\eqref{eq:semiglory} is related to the rainbow scattering. Notably, this semiclassical formula which describes the glory scattering diverges if the rainbow angle is precisely at $\chi=\pi$, due to the divergence of $db/d\chi$. This points to a limitation in the analytical semiclassical description of the backward scattering when the rainbow angle is at $\pi$. We shall show, by using the full numerical partial wave analysis, that there is no actual divergence in such cases, although the amplitude of the differential scattering cross section of the backscattered wave is indeed amplified for large rainbow angles (see Sec.~\ref{sec:semiclassical}).

\section{Numerical results}\label{sec:num_results}

We numerically integrate the radial differential equation, given by Eq.~\eqref{eq:kgr}, from the event horizon, assuming a purely ingoing wave [cf. Eq.~\eqref{eq:inmodes}], up to large distances ($r\gg M$ and $r\gg \omega^{-1}$). By comparing the numerical solution with the expected behavior at the spatial infinity region, we extract the transmission and reflection coefficients. [For more details about the numerical integration of Eq.~\eqref{eq:kgr}, see Ref.~\cite{Macedo2016}.] The convergence of the summation in Eq.~\eqref{eq:scattering} is usually very poor, due to the Coulomb-like characteristics of the potential at long distances. However, we can overcome this problem by implementing a different approach called \textit{reduced series}, introduced in Ref.~\cite{Yennie_1954:prd95_500}. By using a fourth-order reduced series, we find that adding up to $l\sim50$ is enough for the range of frequencies we analyzed.
Our numerical results are stable against changes in the numerical infinity and event horizon, as well as for the maximum value of $l$.

In this section, we present a selection of our numerical results for the differential scattering cross section, considering various DBH configurations---that is, different choices of $\mh$ and $\rs$.

\subsection{Partial wave analysis}

\begin{figure}[h!]%
\includegraphics[width=\columnwidth]{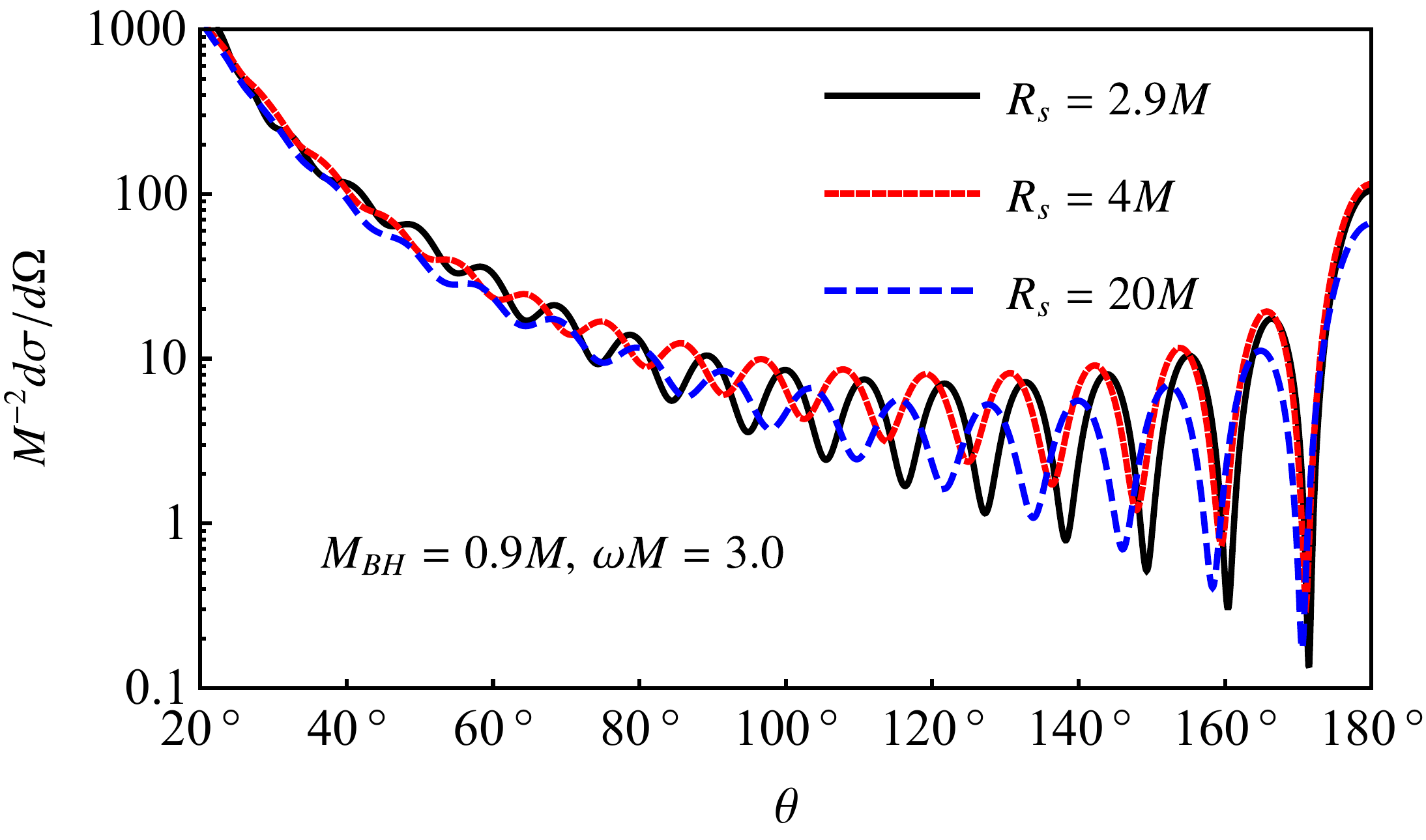}
\caption{Scattering cross section of DBHs for different shell radii, with fixed $\mh=0.9M$ and $\omega M=3.0$. We see that both the interference fringes and the scattering amplitude change with the variation of the radial position of the shell.}%
\label{fig:scatteringfixedmh}%
\end{figure}
Let us focus on the case of a DBH with fixed relative mass $\mh=0.9M$, varying the radial position of the shell. In Fig.~\ref{fig:scatteringfixedmh} we plot the results for scattered waves with frequency~$\omega M=3.0$. The numerical results show that both the amplitude and the interference fringes change when the shell's position is modified. 
For larger values of the shell's radius, the number of interference fringes of the scattering cross section decreases. This happens because the case of large $R_s$ tends to an ISBH with $M=M_{\rm BH}$ (cf. Sec.~\ref{sec:DBHxIBHs}). The scattering amplitude also changes because it scales with the mass of the BH.

\begin{figure}[h!]%
\includegraphics[width=\columnwidth]{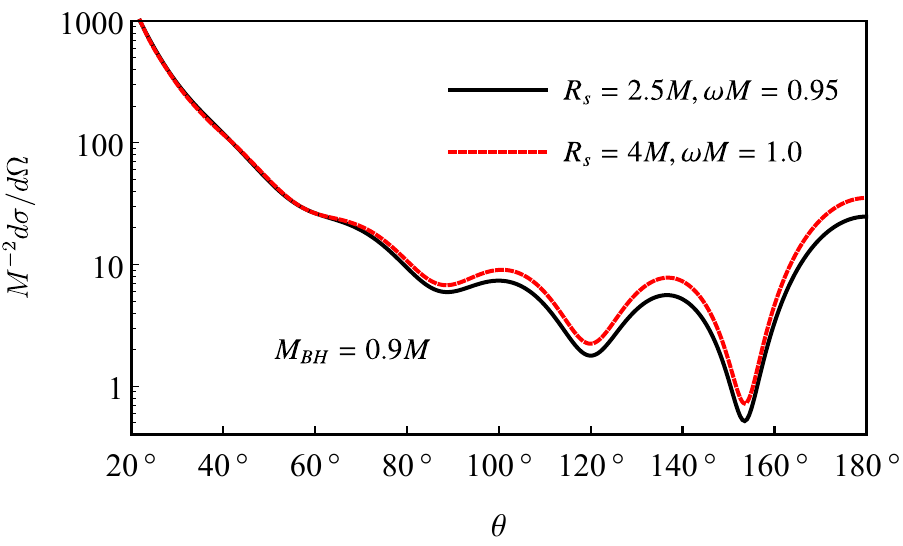}%
\caption{Scattering cross section of different DBH configurations with a similar interference pattern.}%
\label{fig:samepattern}%
\end{figure}
One may ask if different DBH configurations can present similar scattering patterns. This may happen for IBHs and was identified in others contexts~\cite{Macedo:2015qma,Sporea2017}. Interestingly, the scattering of scalar waves with different frequencies can present a similar interference pattern for different DBH configurations. This is illustrated in Fig.~\ref{fig:samepattern}, where we plot two cases with different shell positions and the same $\mh$, exhibiting similar interference patterns.

\begin{figure}[h!]%
\includegraphics[width=\columnwidth]{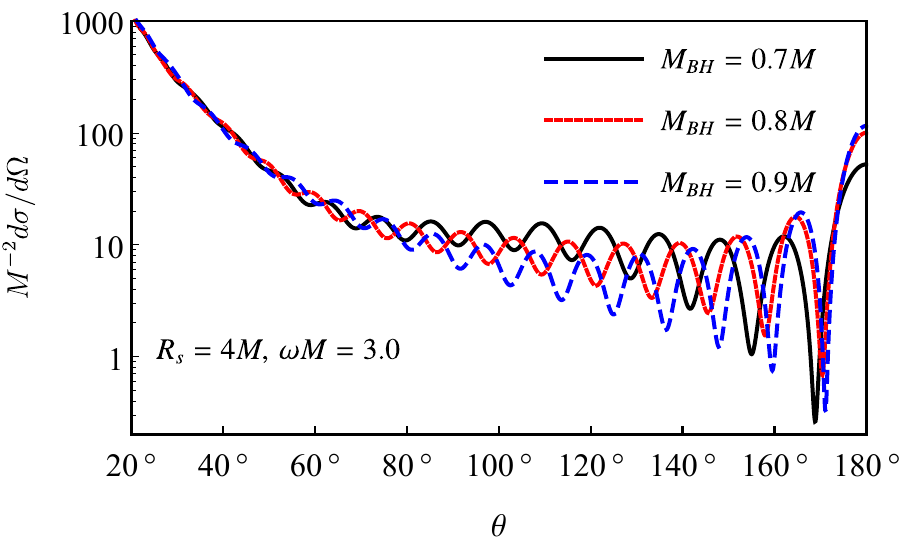}%
\caption{Scattering cross section for a fixed shell's radial position~($\rs=4M$) and different values of $\mh$. All the curves were generated for $\omega M=3.0$. The number of interference fringes increases for higher values of the ratio $\mh/M$.}%
\label{fig:different_mh}%
\end{figure}

It is also important to analyze how the scattering cross section changes for different values of $\mh$, and consequently, different shells' masses. Some results are shown in Fig.~\ref{fig:different_mh}, where we fix the radial position of the shell at $\rs=4M$, changing the value of $\mh$. Note that we are comparing curves with the same wave frequency ($\omega M=3.0$). We see that the decreasing of $\mh$ increases the number of fringes in the scattering pattern. For ISBHs, the number of interference fringes is higher for higher values of the wave frequency, as well as for higher values of the BH's mass. 

\subsection{Classical analysis: Rainbow scattering}
\begin{figure}[h!]%
	\includegraphics[width=\columnwidth]{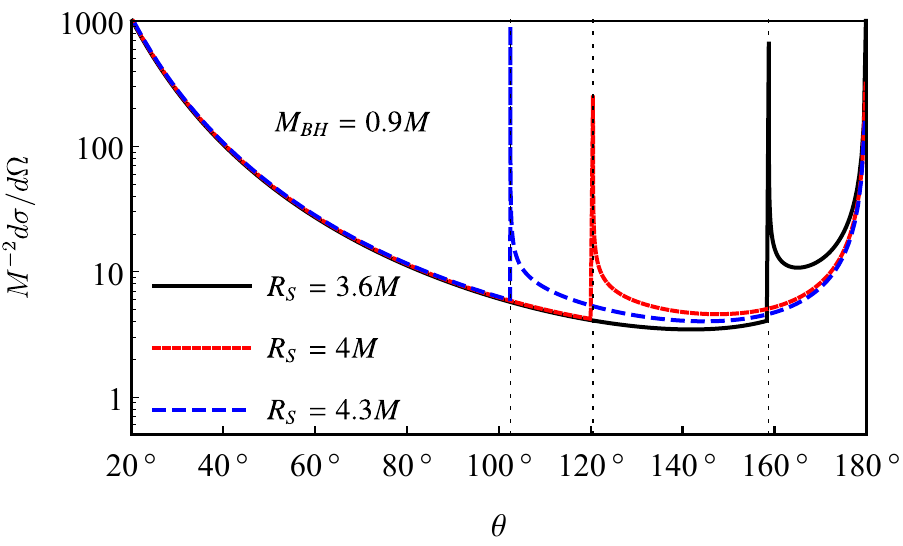}%
	\caption{Geodesic scattering cross section of DBHs with relative mass $\mh=0.9M$ and different positions for the shell $\rs=3.6M$, $4M$, and $4.3M$. In all the cases, there are divergences associated with rainbow angles, located at $\theta_r\approx158.72^{\circ}$~(for $\rs=3.6M$), at $\theta_r\approx120.46^{\circ}$~(for $\rs=4M$), and at $\theta_r\approx102.49^{\circ}$~(for $\rs=4.3M$).}%
	\label{fig:clas_scatt}%
\end{figure}

It has been inferred from the behavior of the deflection angle (cf. Fig.~\ref{fig:geodefleang}) that rainbow scattering can occur in DBH backgrounds. In Fig.~\ref{fig:clas_scatt} we show the geodesic scattering cross section for different DBH configurations. We note that, in all the cases exhibited, the cross sections present divergences related to rainbow angles. Moreover, for this fixed BH relative mass~($\mh=0.9M$), it is possible to see that the closer the shell is located to the BH, the larger the rainbow angle (cf. also Fig.~\ref{fig:geodefleang}).

\subsection{Glory scattering and rainbows}\label{sec:semiclassical}

As mentioned in Sec.~\ref{sec:glory}, the semiclassical glory scattering approach, given by Eq.~\eqref{eq:semiglory}, points to a divergence in the backward scattering whenever the rainbow angle is located exactly at $\theta_r=\pi$. Just like the classical divergence for large angles, we shall show here that the {\it{glory rainbow}}~\footnote{By {\it{glory rainbow}}, we mean $\theta_r\approx\pi$.} divergence is eliminated once the full partial wave analysis is taken into account.

\begin{figure}[h!]%
\includegraphics[width=\columnwidth]{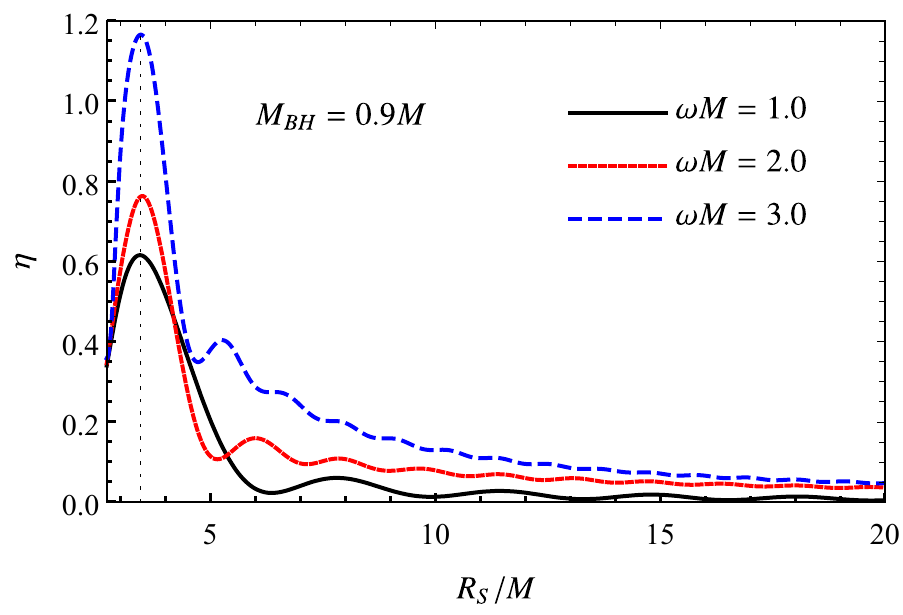}%
\caption{Deviation $\eta$, defined by Eq.~\eqref{eq:deviation}, computed numerically for different frequencies~($\omega M=1.0$, $2.0$, and $3.0$) and $\mh=0.9M$, as a function of the shell's radius. The vertical dotted line locates the shell's position for which the rainbow angle is at $\chi=\pi$, where we see an enhancement in $\eta$, but no divergence. Asymptotically, the DBH scattering cross section tends to the case of an ISBH with mass $\mh$, in a power-law falloff that depends on the frequency of the scattered wave.}%
\label{fig:rainbowglory}%
\end{figure}

In order to analyze the glory rainbow scattering, we define the deviation from the isolated Schwarzschild BH as
\be
\eta\equiv\l.\frac{d\sigma_{\rm DBH}/d\Omega}{d\sigma_{\rm IBH}/d\Omega}\r|_{\theta= \pi}-1,\label{eq:deviation}
\ee
where $d\sigma_{\rm DBH}/d\Omega$ and $d\sigma_{\rm IBH}/d\Omega$ are the differential scattering cross sections for the DBH and for the ISBH with mass $\mh$, respectively. As a representative case, we use $\mh=0.9M$, varying the shell's position. When the shell is located at $\rs=3.45M$ (or $\rs=3.5M$), the rainbow angle $\theta_{r,min}$ (or $\theta_{r,max}$) is located near $180^{\circ}$~(see Fig.~\ref{fig:rainbowangle}), which leads to a divergence in the semiclassical approach to the glory scattering, given by Eq.~\eqref{eq:semiglory}. In Fig.~\ref{fig:rainbowglory}, we plot $\eta$ as a function of the shell's position, computed numerically using the partial wave analysis, choosing different frequencies. We note that the glory rainbow scattering presents a significant increase in the backward direction. This enhancement is improved for high frequencies, which is consistent with the fact that the semiclassical approach is mostly valid in that regime. Additionally, we see that the $\eta$ goes to zero as $\rs$ increases,  meaning that the maximum amplification occurs when the rainbow angle is at $\theta_r=\pi$. Moreover, the differential scattering cross section for the DBH tends to the ISBH case in the large-$R_s$ limit. This is expected, mainly because the energy density of the shell decreases as we increase its radial position~\cite{macedo2018absorption}.

\begin{figure}[h!]%
\includegraphics[width=\columnwidth]{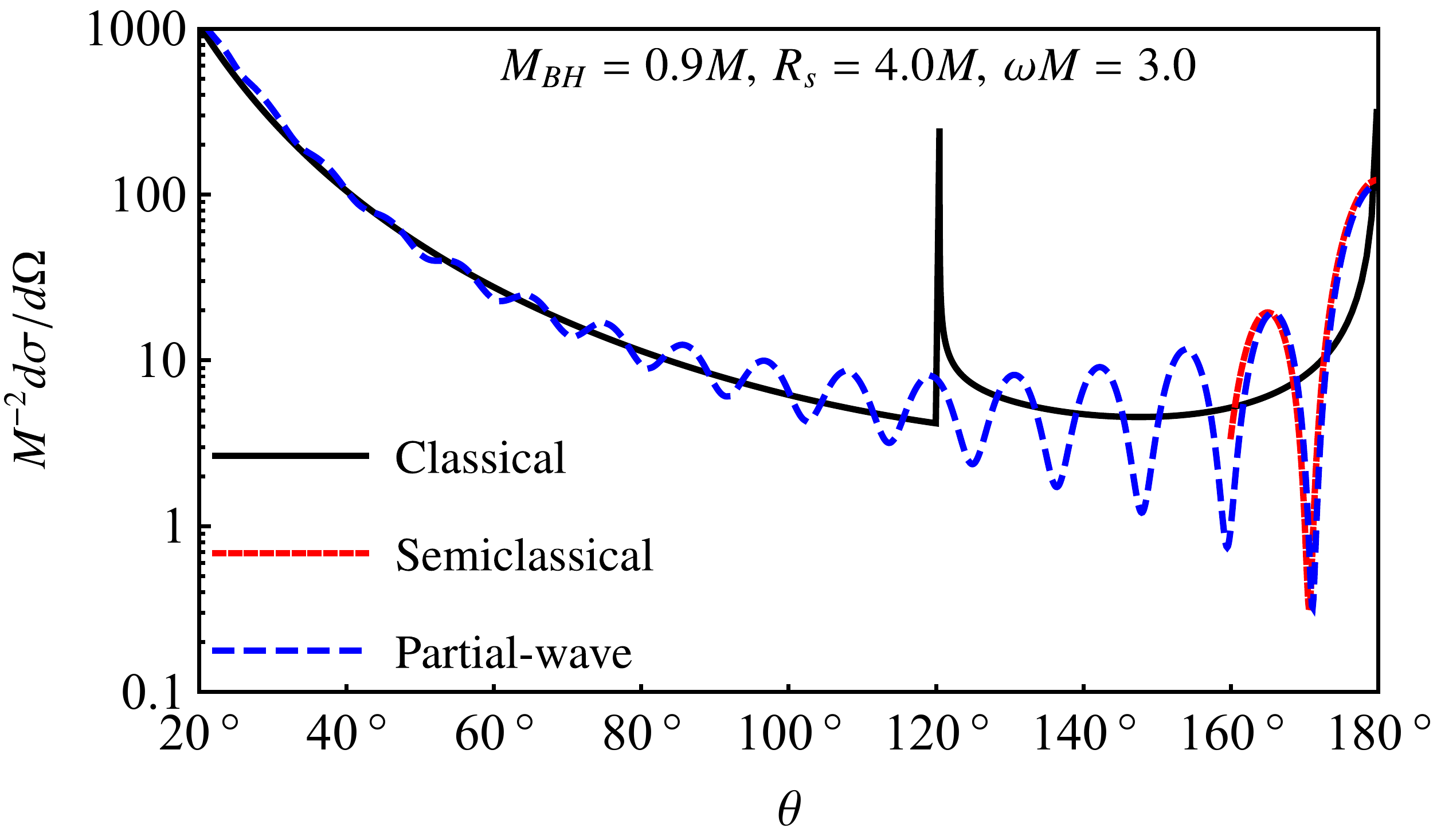}%
\caption{Classical scattering cross section, given by Eq.~\eqref{eq:scatteringgeo}, semiclassical analytical formula for the glory scattering, given by Eq.~\eqref{eq:semiglory}, and numerical results obtained through Eq.~\eqref{eq:scattering}. We have chosen a DBH configuration with $\mh=0.9M$ and $\rs=4.0M$. The numerical and semiclassical results were obtained for $\omega M=3.0$.}%
\label{fig:partialwave_glory}%
\end{figure}

In Fig.~\ref{fig:partialwave_glory}, we exhibit the scattering cross sections for a DBH~(with $\mh=0.9M$ and $\rs=4.0M$) computed numerically, together with the classical scattering cross section, given by Eq.~\eqref{eq:scatteringgeo}, and the semiclassical approach for the glory scattering, given by Eq.~\eqref{eq:semiglory}. We see that at large scattering angles---and away from the cases with $\theta_r\approx \pi$---the semiclassical result for the scattering cross section is in excellent agreement with the numerical result, and at small scattering angles the classical and the partial waves results diverge in the same way, as expected. We have chosen $\omega M=3.0$ for both the semiclassical and the numerical computations.

\subsection{DBHs versus ISBHs}
\label{sec:DBHxIBHs}

\begin{figure}[h!]
	\includegraphics[width=\columnwidth]{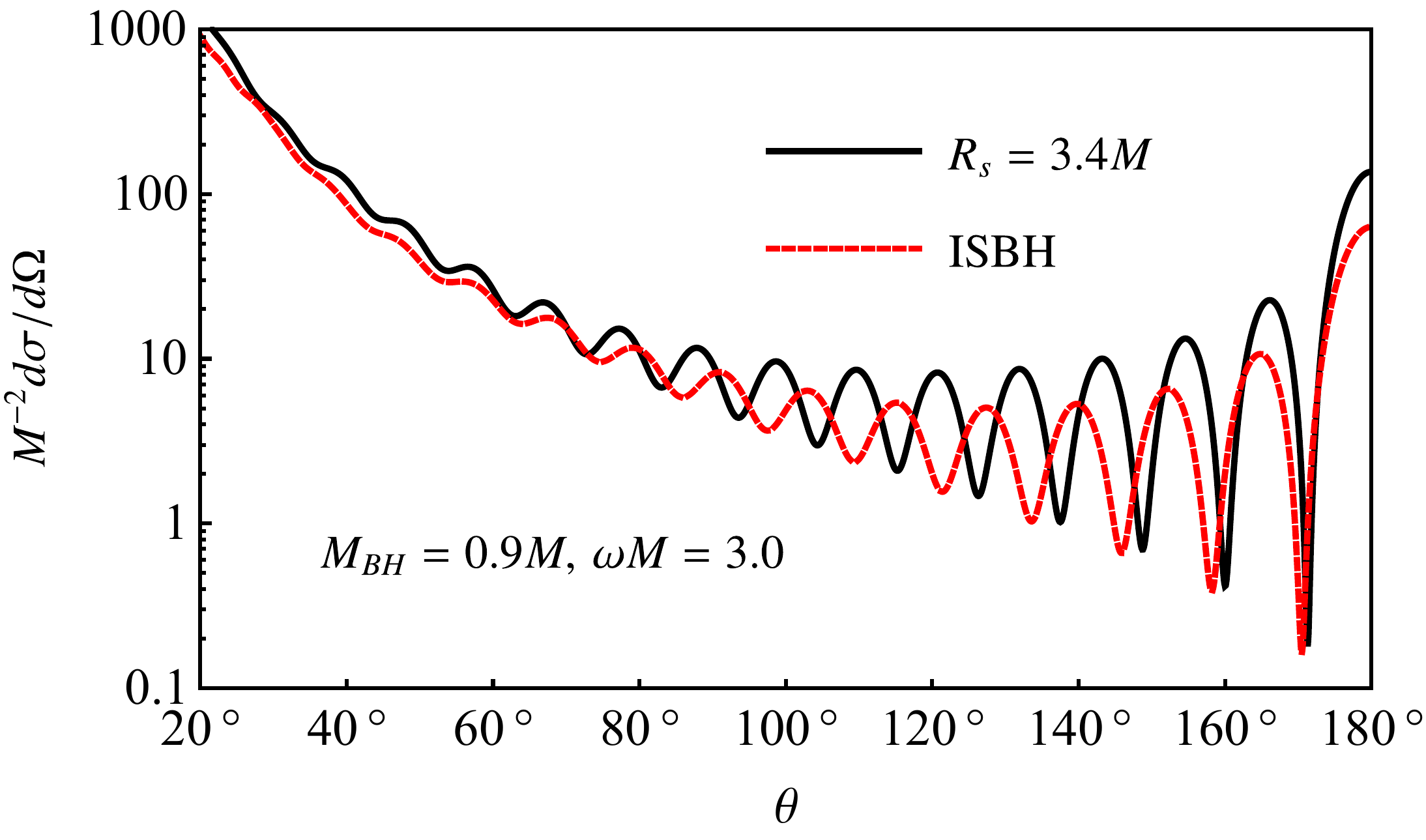}\\
	\caption{Scattering cross section for an ISBH compared with the one for a DBH of same mass~($\mh=0.9M$) and $\rs=3.4M$, the same configuration chosen in Fig.~\ref{fig:geoscat}. We note an enhancement of the glory maximum for this configuration with a large rainbow angle.}%
	\label{fig:scatteringcomp_mh09}%
\end{figure}

Let us now compare the results for DBHs with those for ISBHs. In Fig.~\ref{fig:scatteringcomp_mh09}, we show the scattering cross section for an ISBH together with the one for a DBH of the same mass~($\mh=0.9M$). We choose the same DBH configuration as in Fig.~\ref{fig:geoscat}~(with $\rs=3.4M$), which presents rainbow scattering. The peaks of the DBH orbiting oscillations as well as the backscattered glory are larger than the ISBH ones.

\begin{figure*}[h!]
		\includegraphics[width=\columnwidth]{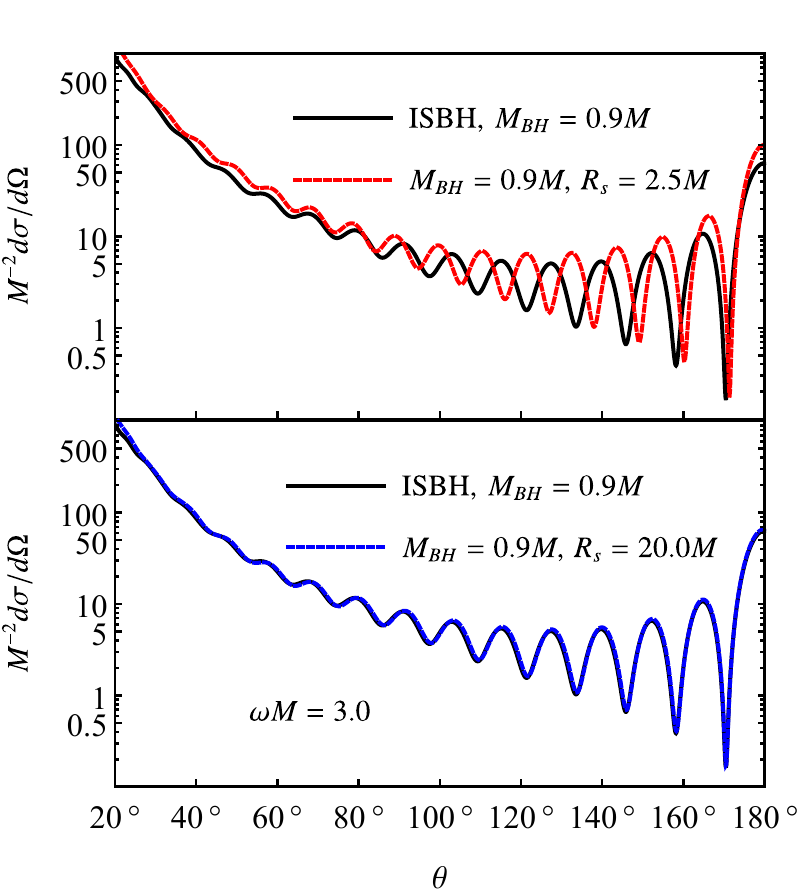}\includegraphics[width=\columnwidth]{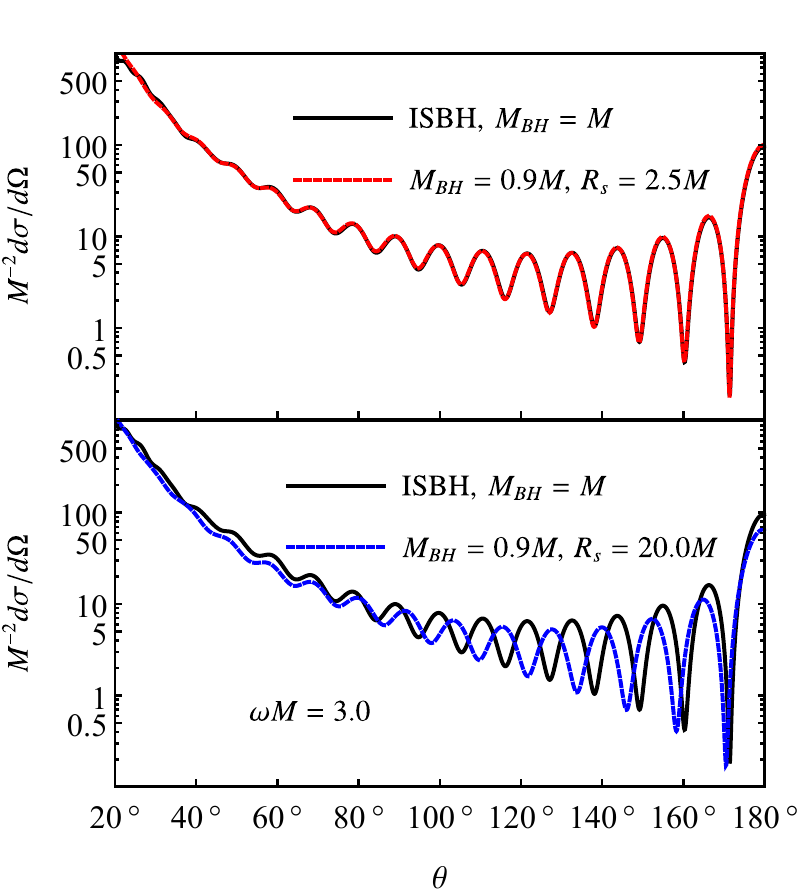}
		\caption{Comparison between scattering cross sections of DBHs and ISBHs. In the left panel, we consider an ISBH with $\mh=0.9M$ and different DBH configurations~[with $\mh=0.9M$ and $\rs=2.5M$ (top) or $20.0M$ (bottom)]. In the right panel, we consider the same DBH configurations used in the left panel and an ISBH with $\mh=M$. Depending on the values of the DBH parameters~($\rs$ and $\mh$), the results of DBHs approach those of ISBHs.}%
		\label{fig:scatteringcomp}%
\end{figure*}

In Fig.~\ref{fig:scatteringcomp}, we exhibit additional results of DBHs and ISBHs for $\omega M=3.0$. We note that in the limits $\rs\to2M$ and $\rs\gg2M$, the fringes have precisely the same patterns as ISBHs with masses $M$ and $0.9M$, respectively.

\section{Final remarks}\label{sec:final_remarks}
We have computed the scattering cross section of planar massless scalar waves for spherically symmetric black holes surrounded by a thin spherical shell, showing that the presence of matter surrounding the BH modifies its scattering properties. Our analysis accounts for changes in the radial position of the shell, as well for changes in the mass fraction of the system. We have shown that in the presence of surrounding matter, the differential scattering cross section presents the same general characteristics of the scattering of planar waves by ISBHs~\cite{anderson2002scattering}---i.e., a divergence in the forward direction, an oscillatory pattern for intermediate-to-large angles---due to the spiral scattering, and a glory peak in the backward direction.

Some nonexistent phenomena in the case of IBHs appear in the case of BHs surrounded by shells. One of them, notably, is rainbow scattering. Rainbow scattering has recently been reported for the case of horizonless compact objects~\cite{Dolan:2017rtj}, and here we have shown that it can happen for BHs when they are surrounded by matter. When the rainbow angle is located at the antipodal direction, we have shown that the divergence presented by the semiclassical glory scattering approximation is associated with an enhaced glory peak, obtained in the context of full numerical partial wave analysis. Furthermore, the rainbow scattering can significantly enhance the backscattered amplitudes, for certain values of the radial position and the mass of the shell.

We have shown that when the shell is located close to the BH, the scattering cross section results approach those for an ISBH with mass equivalent to the total DBH configuration mass. When the shell is placed far away from the BH, the results tend towards those of an ISBH of mass equal to the BH without the thin spherical shell. This is in agreement with the behavior presented by the absorption cross section of DBHs~\cite{Macedo2016}.

We expect the features studied here 
to be present in more realistic scenarios. 
The DBH model adopted in this paper is a simplification, since astrophysical BHs generally present rotation and nonspherical accretion disks. 
To consider more realistic models is a natural extension of our work, allowing a wider understanding of the scattering by DBHs. 

\begin{acknowledgments}
The authors would like to thank Conselho Nacional de Desenvolvimento Cient\'ifico e Tecnol\'ogico (CNPq) and Coordena\c{c}\~ao de Aperfei\c{c}oamento de Pessoal de N\'ivel Superior (CAPES)---Finance Code 001, from Brazil, for partial financial support.
This work has also been supported by the  European  Union's  Horizon  2020  research  and  innovation  (RISE) program H2020-MSCA-RISE-2017 Grant No.~FunFiCO-777740.
\end{acknowledgments}
%
\bibliography{refs}

\end{document}